\documentclass[
superscriptaddress,
preprint,
amsmath,amssymb,
aip, 
article
]{revtex4-2}
\usepackage{graphicx}
\usepackage{dcolumn}
\usepackage{gensymb}
\usepackage{xcolor}
\usepackage{hyperref}
\usepackage{float}
\usepackage{outlines}
\hypersetup{colorlinks=true,linkcolor=black,citecolor=black,urlcolor=black,filecolor=black,pdfborder={0 0 0},}
\usepackage{enumitem}
\usepackage{upgreek}
\usepackage[normalem]{ulem}
\usepackage{color,soul} 

\usepackage{xr}
\makeatletter
\newcommand*{\addFileDependency}[1]{
  \typeout{(#1)}
  \@addtofilelist{#1}
  \IfFileExists{#1}{}{\typeout{No file #1.}}
}
\makeatother

\makeatletter
\renewcommand{\@fnsymbol}[1]{%
  \ifcase#1\or *\or \dagger\or \ddagger\or
  \mathsection\or \mathparagraph\or \|\or **\or
  \dagger\dagger\or \ddagger\ddagger \else\@ctrerr\fi}
\makeatother

\newcommand*{\myexternaldocument}[1]{
    \externaldocument{#1}
    \addFileDependency{#1.tex}
    \addFileDependency{#1.aux}
}
\usepackage{xr}
\makeatletter

\myexternaldocument{SI}

\begin{document}

\noindent\textbf{Integrated polarization-entangled photon source for wavelength-multiplexed quantum networks}

\begin{flushleft}
Xiaodong Shi$^{1,2,3,\#}$, Yue Li$^{4,\#}$, Jinyi Du$^{3,\#}$, Lin Zhou$^{4}$, Ran Yang$^{4,5}$, En Teng Lim$^{3,5}$,\\
Sakthi Sanjeev Mohanraj$^{4}$, Mengyao Zhao$^{4}$, Xu Chen$^{4}$, Xiaojie Wang$^{4}$, Guangxing Wu$^{4}$,\\
Hao Hao$^{3}$, Veerendra Dhyani$^{1,2}$, Sihao Wang$^{1,2}$, Alexander Ling$^{3,5,*}$, and Di Zhu$^{1,2,3,4,*}$
\end{flushleft}

\noindent$^{1}$\textit{A$^\ast$STAR Quantum Innovation Centre (Q.InC), Agency for Science, Technology and Research (A$^\ast$STAR), Singapore 138634, Singapore}\\
$^{2}$\textit{Institute of Materials Research and Engineering (IMRE), Agency for Science, Technology and Research (A$^\ast$STAR), Singapore 138634, Singapore}\\
$^{3}$\textit{Centre for Quantum Technologies, National University of Singapore, Singapore 117543, Singapore}\\
$^{4}$\textit{Department of Materials Science and Engineering, National University of Singapore, Singapore 117575, Singapore}\\
$^{5}$\textit{Department of Physics, National University of Singapore, Singapore 117551, Singapore}

\vspace{0.8em}

{\footnotesize
\noindent\#These authors contributed equally to this work.\\
\noindent *Email: Di Zhu (dizhu@nus.edu.sg), Alexander Ling (alexander.ling@nus.edu.sg)
}

\vspace{1.2em}
\noindent\textbf{ABSTRACT}
\vspace{1.2em}\newline
\noindent Entangled photons are fundamental resources for quantum communication, computing, and networking. 
Among them, polarization-entangled photon pairs play an important role due to their straightforward state manipulation and direct use in quantum key distribution, teleportation, and network protocols. 
However, realizing compact, efficient, and scalable polarization-entangled sources that meet the requirements of practical deployment remains a major challenge.
Here, we present a simple yet high-performance on-chip polarization-entangled photon-pair source on thin-film lithium niobate (TFLN). 
Our device employs dual quasi-phase matching (D-QPM) that sequentially supports type-0 and type-I spontaneous parametric down-conversion in a single nanophotonic waveguide, eliminating the need for interferometers, polarization rotators, or other complex circuits. 
The source directly produces high-fidelity Bell states with broad bandwidth, high brightness, and low noise.
Using this integrated approach, we realize wavelength-multiplexed entanglement distribution in a four-user quantum network deployed over a loop-back metropolitan fiber links up to 50 km.
\vfill

\clearpage

\section*{Introduction}

Entanglement, a uniquely nonclassical correlation between quantum particles, is a cornerstone of quantum information science and underpins applications ranging from secure communication to quantum computing \cite{horodecki2009quantum}. 
Among various forms of entanglement, polarization entanglement plays a central role due to its straightforward state manipulation for direct applications in emerging quantum technologies, such as machine learning on a quantum computer, device-independent quantum key distribution, quantum teleportation, and quantum-enhanced bioimaging \cite{cai2015entanglement,zhang2024entanglement,wengerowsky2018entanglement,yin2020entanglement,bouwmeester1997experimental,yin2017satellite, craddock2024automated, wengerowsky2019entanglement,ursin2007entanglement,yin2012quantum,ma2012quantum,chen2021integrated,zhang2024quantum, yadav2025telecom}.

Polarization-entangled photon pairs can be generated through either type-II or type-0/I nonlinear interactions \cite{anwar2021entangled}.
In type-II spontaneous parametric down-conversion (SPDC), the orthogonal polarizations of signal and idler photons usually have limited phase-matching bandwidth due to polarization-dependent dispersion \cite{kwiat1995new,trojek2004compact,sun2019compact,martin2010polarization}.
Traditional type-0/I SPDC in bulk nonlinear crystals avoids these drawbacks, but requires rotation of the pump or the photon-pair polarization, or crystal orientation, often implemented with delicate free-space optical setups \cite{kwiat1999ultrabright,jabir2017robust,terashima2018quantum,lohrmann2020broadband}. 
Such bulk-optics architectures, though historically successful, are limited in scalability and not suitable for dense integration.

Recent advances in integrated photonics have sought to overcome these limitations \cite{orieux2017semiconductor,signorini2020chip}.
Polarization-entangled photon-pair sources have been demonstrated in silicon and silicon nitride platforms via spontaneous four-wave mixing, and in thin-film lithium niobate (TFLN) via type-0 SPDC \cite{jiang2025entanglement,du2024demonstration,matsuda2012monolithically,lv20131,olislager2013silicon,zhang2024polarization,hua2025bright,kim2025integrated,jiao2025electrically}. 
These implementations typically rely on auxiliary polarization-handling components, such as polarization beam splitters, polarization rotators, two-dimensional gratings, and cross-polarized dual cavities, which may increase system complexity and constrain operational bandwidth. 
AlGaAs platform exploits both $\chi^{(2)}$ and $\chi^{(3)}$ nonlinearities to realize concurrent type-0 and type-I entanglement generation with high brightness, but they lack flexible and independent control over the two processes \cite{kultavewuti2017polarization,appas2021flexible,kang2015two}.
Thus, despite significant progress, there remains strong interest in developing integrated polarization-entangled photon sources that are high-performance, controllable, and architecturally simple.

In this work, we introduce a dual quasi-phase matched (D-QPM) periodically poled lithium niobate (PPLN) nanophotonic waveguide, that directly generates polarization-entangled photon pairs via sequential type-0 and type-I SPDC in a single device. 
This design eliminates the need for interferometric or polarization-manipulating circuits, while offering flexible and in situ phase-matching control and phase tuning.
The source exhibits high brightness, high fidelity, and broadband operation, enabling wavelength-multiplexed entanglement distributions.
We further demonstrate long-distance, four-user quantum networking over deployed metropolitan fiber system, underscoring its practicality for real-world quantum communications. 
Our work establishes a simple, compact, robust, and manufacturable pathway for on-chip polarization-entangled photon-pair source, marking an important step toward scalable quantum mesh networks and introducing a viable integrated approach to Bell-state generation.

\section*{Results}

\subsection*{Device principle}

\begin{figure*}[htbp]
\centering 
\includegraphics[width = 6.3in]{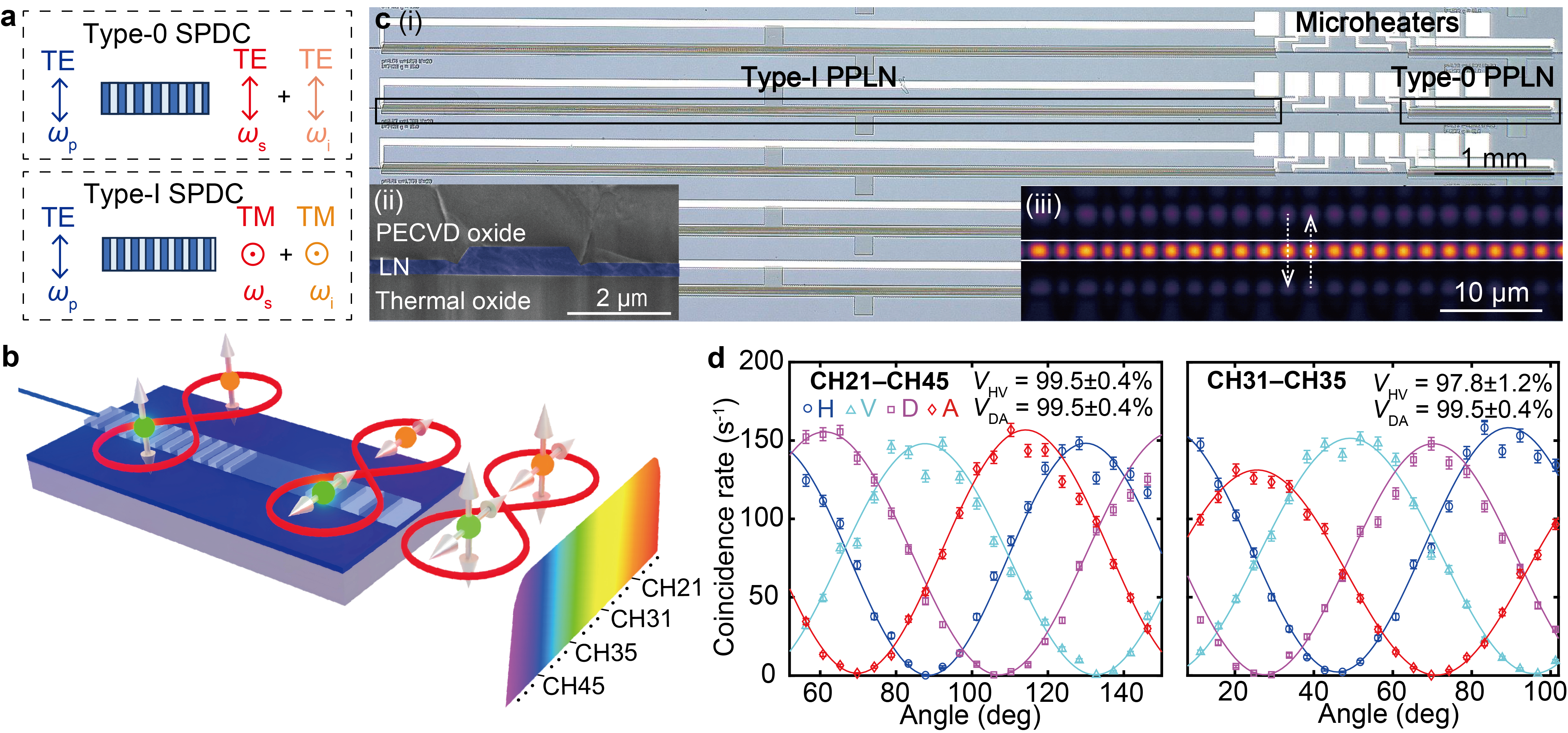}
\caption{\textbf{Dual quasi-phase matching (D-QPM) polarization-entangled photon-pair source in thin-film lithium niobate.} \textbf{a} Schematics of type-0 and type-I SPDC in PPLN waveguides with distinct poling periods for phase matching. In type-0 SPDC, a TE pump photon converts to a pair of TE signal and idler photons, while in type-I SPDC, a TE pump photon converts to a pair of TM signal and idler photons. \textbf{b}, Schematic of wavelength-multiplexed polarization-entangled photon-pair source. The device is basically a single straight waveguide incorporating two PPLN sections with distinct poling periods, enabling orthogonally polarized photon-pair generation via type-0 and type-I SPDC. \textbf{c}, (i) Optical micrograph of fabricated D-QPM PPLN waveguide array. Each device integrates two PPLN sections for type-0 and type-I SPDC and three microheaters for independent tuning of their phase-matching wavelengths and relative phase. (ii) False-color cross-sectional scanning electron micrograph of the waveguide. (iii) Laser-scanning SHG imaging of the type-I poled region. \textbf{d}, Two-fold coincidence measurement in H, V, D, and A bases for far non-degenerate (CH21-CH45) and near-degenerate (CH31-CH35) photon pairs. The error bars are estimated by Poissonian photon-counting statistics. Raw visibilities are measured to be $>$97$\%$ without accidental subtraction, confirming high-quality polarization-entangled photon-pair source.}
\label{Fig1}
\end{figure*}

The D-QPM PPLN nanophotonic waveguide for producing polarization-entangled photon pairs is designed on 600 nm thick $x$-cut TFLN.
Two types of phase matching (type-0 and type-I) can be realized in a single PPLN waveguide with different poling periods (Fig.\ref{Fig1}a).
Similar concept has been theoretically proposed in potassium titanyl phosphate (PPKTP) crystal \cite{shukla2020generation}.
In type-0 SPDC, a transverse-electric (TE) pump photon converts to a pair of TE signal and idler photons, exploiting the $\chi ^{(2)}$ nonlinear coefficient $d_{33}$ \cite{zhu2021integrated}.
While in type-I SPDC, a TE pump photon converts to a pair of transverse-magnetic (TM) signal and idler photons, leveraging $d_{31}$.
The core device to produce polarization-entangled photon pairs on TFLN is basically a single straight LN nanophotonic waveguide comprising two sequentially arranged sections of PPLN, each engineered with a distinct poling period (Fig.\ref{Fig1}b).
One PPLN section is designed for type-0 quasi-phase matching (QPM) between the fundamental-harmonic (FH) TE mode and the second-harmonic (SH) TE mode.
The other section supports type-I QPM between the FH TM mode and SH TE mode.

When the dual-QPM PPLN device is pumped at the SH wavelength with TE polarization, both type-0 and type-I SPDC processes occur sequentially, generating TE-polarized ($\mathrm{|H_sH_i\rangle}$) and TM-polarized ($\mathrm{|V_sV_i\rangle}$) photon pairs around the FH wavelength, respectively.
The resulting polarization-entangled state at the waveguide output can be expressed as $|\psi\rangle = \alpha\mathrm{|H_sH_i\rangle} + \beta e^{i\phi}\mathrm{|V_sV_i\rangle}$,
where $\alpha$ and $\beta$ are the probability amplitudes of the two SPDC processes, and $\phi$ represents their relative phase.
Maximal entanglement is obtained when $\alpha = \beta$ and $\phi = 0$ or $\pi$, yielding $|\psi_{max}\rangle = \frac{1}{\sqrt{2}}(\mathrm{|H_sH_i\rangle}\pm\mathrm{|V_sV_i\rangle})$.
To achieve the conditions, we tailor the lengths of each PPLN section to balance pair generation rates and introduce a thermo-optic phase shifter between them to fine-tune the relative phase.

The D-QPM PPLN devices are fabricated (Fig.\ref{Fig1}c, see Methods for fabrication details).
The type-0 PPLN section is 1.20 mm long with a poling period of 4.76 \textmu m, while the type-I section is 7.43 mm long with a poling period of 4.54 \textmu m.
The lengths are designed according to their nonlinear strengths to ensure identical pair generation rates, and the poling periods are calculated based on type-0/I QPM conditions (see Methods for design details).
Integrated microheaters, positioned above each PPLN region and in between, enable independent tuning of the QPM wavelengths and relative phase, respectively.

To demonstrate polarization entanglement, we balance the pair generation rates from the two types of SPDC in the D-QPM PPLN device via thermal tuning. 
The generated photons are spectrally separated via external dense wavelength division multiplexing (DWDM) channels.
Two representative wavelength channel pairs are chosen for polarization entanglement measurements: one far non-degenerate pair (CH21–CH45) and one near-degenerate pair (CH31–CH35) (Fig.\ref{Fig1}d). 
The high visibilities indicate that the proposed architecture produces high-fidelity polarization-entangled photon pairs. 

\subsection*{Classical characterization of D-QPM PPLN waveguide}

\begin{figure*}[htbp]
\centering 
\includegraphics[width = 6.3in]{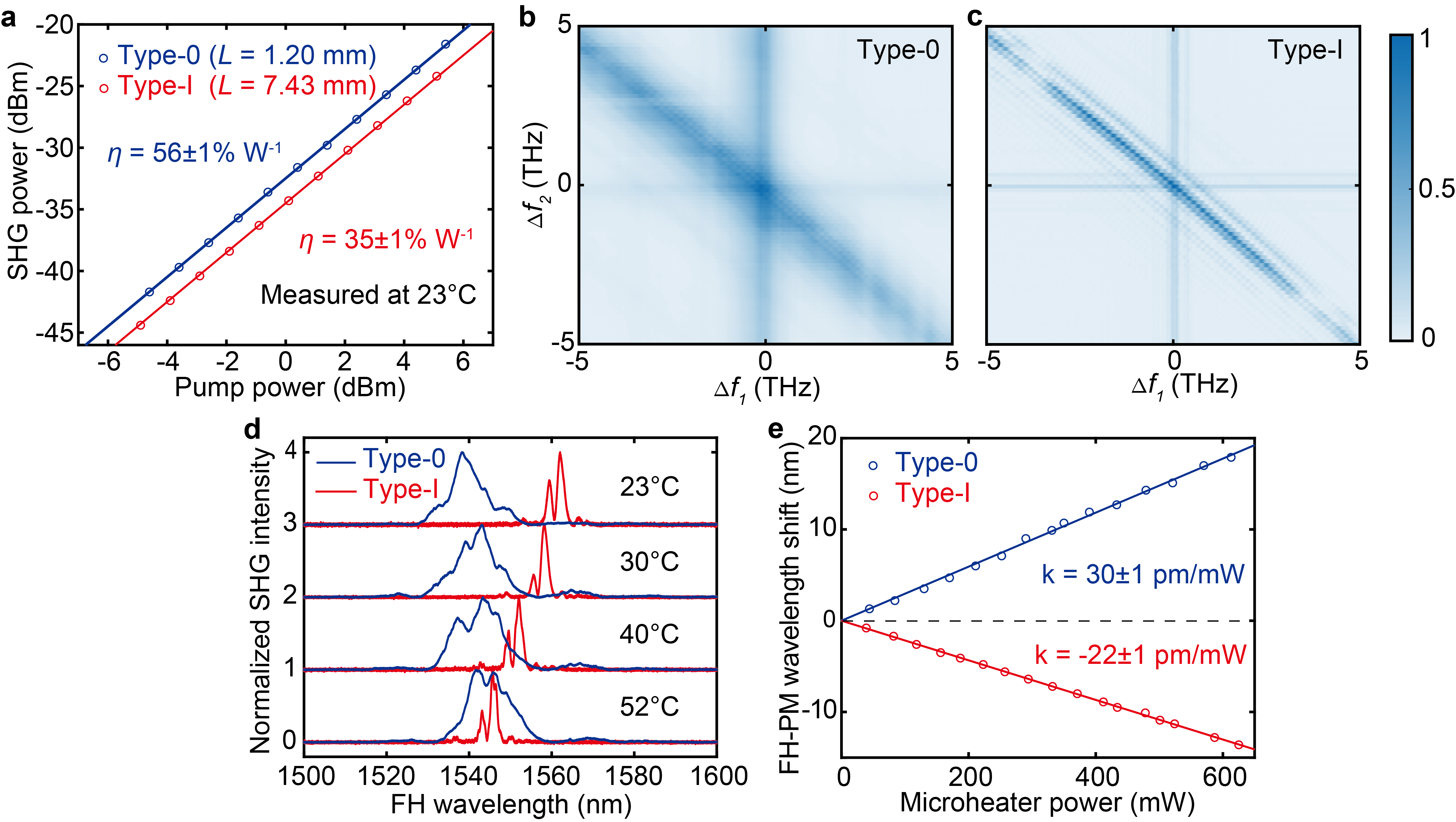}
\caption{\textbf{Classical characterization of D-QPM PPLN waveguide.} \textbf{a}, Measured type-0 (blue circles) and type-I (red circles) on-chip SHG power as a function of pump power, and linear fits (lines) yield conversion efficiencies of 56 $\pm$ 1$\%$ and 35 $\pm$ 1$\%$, respectively. \textbf{b}, \textbf{c}, Sum-frequency generation spectra for type-0 and type-I PPLN waveguides. The anti-diagonal features indicate broadband phase matching for both SPDC processes. \textbf{d}, Type-0 (blue) and type-I (red) SHG spectra at different chip temperatures. The type-0 spectrum red shifts, whereas type-I blue shifts with increasing temperature. The phase-matching wavelength difference between type-0 and type-I PPLNs can be compensated through thermal tuning. \textbf{e}, Fundamental-harmonic (FH) phase-matching wavelength as a function of applied microheater power for type-0 (blue) and type-I (red) SHG. Wavelength shifts of 30 $\pm$ 1 pm mW$^{-1}$ (type-0) and -22 $\pm$ 1 pm mW$^{-1}$ (type-I) demonstrate independently efficient thermal tuning of each PPLN section.}
\label{Fig2}
\end{figure*}

The device's classical performance is evaluated through second-harmonic generation (SHG) and sum-frequency generation (SFG) measurements  (see Supplementary Note 1 for experimental details). 
The SHG conversion efficiencies of the type-0 and type-I PPLN waveguides are obtained by continuous-wave (CW) pumping at their respective phase-matched FH wavelengths with TE and TM polarizations at room temperature, respectively (Fig.\ref{Fig2}a). 
The measured normalized conversion efficiencies are 56.5 $\pm$ 0.8$\%$ W$^{-1}$ for type-0 and 35.5 $\pm$ 0.7$\%$ W$^{-1}$ for type-I, with the difference attributed to fabrication imperfections such as variations in poling depth, duty cycle, and thin-film thickness.
The phase-matching (PM) characteristics are further examined via SFG measurements (Figs.\ref{Fig2}b, c; also see simulations in Supplementary Figure 1).
The anti-diagonal slopes of both PPLN follow the energy conservation line ($\Delta f_1 + \Delta f_2 = 0$), indicative of broadband SPDC. 
The inferred SPDC bandwidths from the SFG measurements are about 78 nm and 51 nm for the type-0 and type-I processes \cite{lenzini2018direct,kaneda2020spectral}.

The deviation in observed PM wavelengths from the design and the non-ideal SHG spectra are mainly attributed to the fabrication imperfections.
To compensate for the PM shifts, we characterize the thermal response of the PPLN PM wavelengths. 
As the chip temperature increases, the type-0 SHG spectrum red shifts, whereas the type-I spectrum blue shifts (Fig.\ref{Fig2}d). 
Thus, alignment of the two PM wavelengths can be achieved by increasing the temperature. 
Individual PM wavelength adjustments are performed by applying electrical power to the integrated microheater on top of each PPLN section. 
The FH PM wavelength shifts by 30 $\pm$ 1 pm mW$^{-1}$ for type-0 and -22 $\pm$ 1 pm mW$^{-1}$ for type-I SHG (Fig.\ref{Fig2}e).
This flexible and independent on-chip thermal control on the PM wavelengths of the two PPLN sections also enables precise balancing of the pair generation rates from the type-0 and type-I SPDC processes, a critical requirement for achieving high-fidelity polarization-entangled states.

\subsection*{Non-classical characterization of D-QPM PPLN waveguide}

We next balance the type-0 and type-I SPDC rates through thermal tuning and characterize their non-classical performances separately before evaluating polarization entanglement (Fig.\ref{Fig3}a, see Supplementary Note 2 for experimental details).
In practice, the pair generation rates from the two SPDC processes have initial difference, due to fabrication imperfections, such as thin-film thickness nonuniformity, waveguide linewidth variation, and waveguide losses. This imbalance can be compensated by thermally tuning the PM conditions of the two PPLN sections with independent microheaters. By adjusting the device temperature independently, the SPDC efficiency of the two processes can be fine tuned to achieve comparable photon-pair brightness within the investigated bandwidth (see details in Supplementary Figure 2).
We measure the pair generation rate and coincidence-to-accidental ratio (CAR) as a function of pump power from a randomly chosen channel pair (CH21-CH45) (Fig.\ref{Fig3}b, c). 
For type-0 SPDC, the measured pair efficiency is $(2.6 \pm 0.1) \times$10$^3$ s$^{-1}$mW$^{-1}$, corresponding to an inferred on-chip efficiency of $2.1 \times 10^7$ s$^{-1}$mW$^{-1}$. 
For type-I SPDC, the measured efficiency is $(2.5 \pm 0.1) \times 10^3$ s$^{-1}$mW$^{-1}$, yielding an on-chip efficiency of $2.0 \times 10^7$ s$^{-1}$mW$^{-1}$. 
Considering the 0.3 nm DWDM channel bandwidth, the source spectral brightness is $\sim 6.7 \times 10^7$ s$^{-1}$mW$^{-1}$nm$^{-1}$. 
The CARs are above $5\times 10^2$ across all measurements and exceed  $5\times 10^3$ at pump powers below 0.03 mW for both SPDC.

Heralded second-order correlations ($g^{(2)}_H(\tau)$) are measured by splitting photons in CH45 and detecting multi-photon events (Fig.\ref{Fig3}a(iv)). 
The measured $g^{(2)}_H(\tau)$ at zero time delay ($g^{(2)}_H(0)$) for type-0 and type-I SPDC are both below 0.01 at 0.63 mW pump power (Fig.\ref{Fig3}d, e), confirming operation deep in the single-photon regime.

Joint spectral intensities (JSI) are reconstructed using 32 DWDM channels (CH17 to CH32 and CH49 to CH34) (Fig.\ref{Fig3}f, g). 
The JSI for type-0 and type-I SPDC shows flat pair generation rates across 16 telecom C-band channel pairs, thanks to the broadband PM conditions of both processes.
The strong frequency correlations over multiple wavelength channels highlight the potential of the source for wavelength-multiplexed quantum applications.

\subsection*{Polarization entanglement measurements}

After balancing and individually characterizing the type-0 and type-I SPDC processes, we verify the polarization entanglement generated by the source. 
The photon pairs (Fig.\ref{Fig3}a) are spectrally separated via DWDM channels (Fig.\ref{Fig4}c).
Two representative channel pairs are randomly chosen for polarization-entanglement measurements: a far non-degenerate pair (CH21–CH45) and a near-degenerate pair (CH31–CH35) (Fig.\ref{Fig1}d), with respect to the degenerate wavelength corresponding to ITU channel CH33, which highlights the broadband operation of the source.
For each channel pair, polarization correlations are measured in horizontal (H), vertical (V), diagonal (D), and anti-diagonal (A) bases (see Supplementary Note 3 for experimental details) \cite{yin2017satellite,kwiat1995new}.
Two-fold coincidence for the two pairs is measured with raw visibilities of $V_{\rm{H/V}} = 99.5 \pm 0.4 \%$ and $V_{\rm{D/A}} = 99.5 \pm 0.4 \%$ for CH21–CH45, and $V_{\rm{H/V}} = 97.8 \pm 1.2 \%$ and $V_{\rm{D/A}} = 99.5 \pm 0.4 \%$ for CH31–CH35, without accidental subtraction. 
The results exhibit sinusoidal curves with clear constructive and destructive interferences, confirming strong quantum correlations between the signal and idler photons.
The corresponding Bell-state fidelities are estimated to be $99.8 \pm 0.1 \%$ and $99.3 \pm 0.3 \%$, close to an ideal Bell state \cite{meyer2018high,weinbrenner2024certifying}.

To demonstrate the source’s suitability for long-distance multi-user quantum networks, entangled photons are distributed over deployed metropolitan optical fibers (provided by National Quantum-Safe Network) connecting four locations across Singapore (labeled as Alice, Bob, Charlie, and David), and looped back for detection (Fig.\ref{Fig4}a-c).
Concurrently using two DWDM channel pairs (CH21–CH45 and CH31–CH35), we configure three combinations (red, blue, and yellow) to realize a four-user quantum network (Fig.\ref{Fig4}d). 
Taking the red combination as an example, CH21–CH45 are distributed to Alice and David paths, while CH31–CH35 are distributed to Bob and Charlie paths, for entanglement distribution between each pair of nodes.
Raw visibilities for each combination, as well as the total link distances, are summarized (Fig.\ref{Fig4}e). 
For instance, when CH21–CH45 are sent to Alice and David (29.2 km) and CH31–CH35 to Bob and Charlie (25.9 km), the measured visibilities are 
$V_{\rm{H/V}} = 93.2 \pm 2.5\%$, $V_{\rm{D/A}} = 91.3 \pm 2.8\%$ and 
$V_{\rm{H/V}} = 92.1 \pm 2.9\%$, $V_{\rm{D/A}} = 91.0 \pm 3.1\%$, corresponding to fidelities of $96.1 \pm 0.9\%$ and $95.8 \pm 1.1\%$, respectively. 
Similar measurements for the other combinations yield fidelities ranging from 93$\%$ to 98$\%$.
All measured visibilities far exceed the 70.7$\%$ threshold for violating CHSH inequality, and the overall good fidelities indicate the robustness of the source for some practical quantum applications, such as entanglement swapping and entanglement-based quantum key distribution protocols \cite{sun2017entanglement,davis2025entanglement,brunner2014bell,scarani2009security,wengerowsky2020passively,pathumsoot2024boosting}.

\section*{Discussion}

To place our results in context, we compare the key performance metrics of recently reported integrated polarization-entangled photon sources on various platforms in Table \ref{tab:tableS}, including photon-pair generation rate, CAR, and entanglement fidelity. 
For a fair comparison, the pair rates are normalized to a 100-GHz bandwidth, assuming DWDM-based wavelength multiplexing. 
This comparison shows that the source demonstrated here simultaneously achieves high pair rate, high CAR, and high fidelity, underscoring its suitability for scalable and practical quantum-network deployment.

There still remain several avenues for improving both device performances and its applications in quantum networks. 
The relatively low measured photon-pair rates, compared with the intrinsic source brightness, are primarily limited by the insertion loss of the deployed fiber system and the chip-to-fiber coupling loss. 
Temporal measurements (see details in Supplementary Figure 3 and 4) show that occasional drifts of the detected photon-pair rate may occur, owing to chip-to-fiber coupling variations induced by laboratory temperature fluctuations. 
Despite the photon-pair rate drifting, the Bell-state fidelity remains consistently high, demonstrating the robustness of the generated polarization-entangled state against the coupling drifts.
Incorporating more efficient and mechanically stable coupling schemes, such as trident edge couplers or SU8 tapers with high-numerical-aperture fibers, or permanent fiber-chip packaging, could substantially reduce these bottlenecks \cite{shi2025squeezed,liang2022efficient,du2024demonstration1}.
Active monitoring on both types of SPDC pair generation rates together with output pump power would also enable automatic compensation of temperature fluctuations and chip-to-fiber coupling drifts, helping to maintain stable operation of the source throughout long-term measurements.
Alternative integrated architectures based on two type-0 PPLN sources can potentially achieve higher brightness by fully utilizing the largest $d_{33}$ nonlinear coefficient. Such approaches provide another promising route toward integrated polarization-entangled photon sources \cite{kim2025integrated,jiao2025electrically}.
The PM tuning and photon-pair brightness balancing demonstrated in this work could also provide a practical strategy for compensating fabrication variations and maintaining balanced photon-pair generation in such architectures.

The fidelity of entanglement distribution through deployed fiber networks is found to decrease relative to direct source measurements. 
The degradation primarily arises from accumulated polarization-mode dispersion (PMD) in the long-distance fibers, which distorts the polarization states of the entangled photons \cite{gisin2002quantum}.
In our network demonstration, a loopback configuration is used, effectively doubling the optical path length between nodes. 
As a result, both the fiber loss and accumulated PMD are twice those expected in an actual network, and the fidelities reported here represent conservative estimates of real-world performance.
The deployed fiber network itself exhibits good polarization stability over the measurement duration (see details in Supplementary Figure 5).
Nevertheless, implementing active polarization stabilization or advanced compensation strategies would further help to mitigate the PMD effect and maintain high-fidelity transmission over extended distances for long-term operation \cite{rodimin2025impact}.
Alternatively, narrowing the photon-pair spectral bandwidth, for instance, by employing microring resonators that generate ultra-bright, narrow-band photon pairs, can further suppress the impacts of both PMD and chromatic dispersion \cite{miloshevsky2024cmos}.
These approaches will collectively enhance the system robustness and scalability for practical quantum network deployment.

Our current network demonstration distributes entanglement among four nodes using two pairs of wavelength channels. 
Due to the limited number of SNSPDs in the current experimental setup, characterization is performed for one user pair at a time, though the wavelength channels for all pairs are simultaneously present in the system.
Leveraging the broadband operation of our polarization-entangled photon-pair source would allow full utilization of available DWDM channels across the telecom C-band, supporting simultaneous entanglement distribution among more users.
Implementing a fully distributed architecture with independent detectors and synchronized time-stamping systems at each node, rather than the looped-back scheme, is an important next step toward achieving a fully functional quantum mesh network, in which independent users can simultaneously share and verify entanglement across the network \cite{neumann2022continuous}.

In summary, our integrated D-QPM PPLN architecture provides a simple and scalable solution for on-chip generation of Bell states. 
By sequentially implementing type-0 and type-I SPDC, and balancing their contributions via integrated microheaters, the device produces high-fidelity polarization entanglement without the need for external polarization components or interferometric schemes. 
The combination of high brightness, low noise and flexible tunability supports robust multi-channel entanglement distribution and positions this source as a strong candidate for networked quantum technologies.

\section*{Methods}
\subsection*{Device design}
The D-QPM PPLN nanophotonic waveguide is designed along the crystal $y$-axis on a 600 nm thick $x$-cut TFLN integrated platform.
The waveguide has a width of 2 \textmu m with an etching depth of 300 nm.
The poling periods for type-0 and type-I QPM process is calculated to be 4.76 \textmu m and 4.54 \textmu m at room temperature, according to $\varLambda = \frac{\lambda_{\rm{FH}}}{2(n_{\rm{SH}}-n_{\rm{FH}})}$, where $\lambda_{\rm{FH}}$ is the FH wavelength (1550 nm) and $n_{\rm{SH/FH}}$ is the corresponding effective index of the involved SH/FH modes.
We evaluate the nonlinear efficiency of the type-0 and type-I QPM processes based on the SHG conversion efficiency, $\eta = \frac{8 \pi^2}{\epsilon_0 c n^2_{\rm{FH}} n_{\rm{SH}} \lambda_{\rm{FH}} ^2 } d^2_{\rm{eff}}\varGamma L^2$, where $\epsilon_0$ is the free-space permittivity, $c$ is the speed of light, $d_{\rm{eff}}$ is the effective $\chi ^{(2)}$ coefficient ($d_{\rm{eff}} = \frac{2}{\pi}d_{\rm{33}}$ for type-0 and $d_{\rm{eff}} = \frac{2}{\pi}d_{\rm{31}}$ for type-I PM processes), and $L$ is the length of the waveguide \cite{shi2024efficient}.
$\varGamma$ is the nonlinear coupling parameter between the FH and SH modes, given by $\varGamma  = \frac{|\int_{{\rm LN}}{ (E^*_{x/z,{\rm FH}})^2 E_{z,{\rm SH}}{\rm d}x{\rm d}z|^2}}{{|\int_{{\rm all}}|E_{\rm {FH}}|^2 {\rm d}x{\rm d}z|^2} \int_{\rm all}{|E_{\rm SH}|^2{\rm d}x {\rm d}z}}$. 
Here, we only consider the overlap of the $x$ and $z$-components of the electric fields ($E$) for the TM and TE modes, respectively.
To match the nonlinear strength of the type-I and type-0 PM process, the ratio of the corresponding PPLN lengths is $\sim 6.19$, so we choose 7.43 mm and 1.20 mm, respectively, in the experiment.

\subsection*{Device fabrication}

We fabricate the D-QPM PPLN device in a 600 nm thick MgO-doped $x$-cut TFLN chip (NanoLN).
A 15 nm thick HfO$_2$ layer is deposited on TFLN using atomic-layer deposition (ALD) as a buffer layer for electrical poling. 
The comb-like electrode is patterned, using e-beam lithography, e-beam metal evaporation (60 nm Ni and 60 nm Cr) and lift-off.
We apply a series of electrical pulses to periodically reverse the polarity in the LN thin film.
The waveguides are patterned using e-beam lithography and Ar$^+$ etching by inductively coupled plasma reactive ion etching (ICP-RIE), with MaN e-beam resist as the etching mask. 
A 1.6 \textmu m thick silicon dioxide thin film is then deposited on top of the waveguide by plasma-enhanced chemical vapor deposition (PECVD).
The thermal microheater is finally patterned, using ultraviolet lithography, e-beam metal evaporation (90 nm Pt) and lift-off.
Overview of fabricated device array, scanning electron micrograph of waveguide cross section, and top-view laser-scanning SHG imaging of the poled waveguide are shown in Fig.\ref{Fig1}c.

\section*{Data availability}
The data supporting this study are provided in the Source Data file associated with this manuscript (doi.org/10.6084/m9.figshare.32981522).

\bibliography{ref_main}

\section*{Acknowledgements}
The authors thank Jing Yan Haw, Hao Qin, and Matthew Wee for providing access to the deployed fiber infrastructure, Christian Kurtsiefer and Hou Shun Poh for their valuable advice on quantum networks, and Hoi-Kwong Lo for fruitful discussions.
The authors also acknowledge Netlink Trust for provisioning the fiber network used in this work.

\section*{Funding Statement}
This research was supported by National Research Foundation, Singapore (NRF-NRFF15-2023-0005, W24Q3D0001, W24Q3D0003), Singapore Ministry of Education (MOET32024-0009), A*STAR (M23M7c0125), and Centre for Quantum Technologies Funding Initiative (S24Q2d0009).

\section*{Author contributions}
D.Z., A.L., X.S., Y.L., and J.D. conceived the idea. X.S. and Y.L. designed the devices. Y.L., S.S.M., X.C., H.H., and V.D. fabricated the devices. X.S., Y.L., J.D., L.Z., R.Y., M.Z., X.W., G.W., and S.W. performed the classical and non-classical measurements of the devices. J.D., X.S., Y.L., and E.T.L. performed the entanglement and network measurements. D.Z. and A.L. supervised the project. All the authors discussed the results and wrote the manuscript.

\section*{Competing interests}
The authors declare no competing interests.

\clearpage

\begin{table*}[tbh!]
\begin{center}
\caption{Summary of integrated polarization-entangled photon sources, including material platforms, nonlinear processes, pump schemes, operating pump power (a: average power; p: peak power), pair generation rate, coincidence-to-accidental ratio, and fidelity. }\label{tab:tableS}
\begin{tabular}{  w{c}{2cm}  w{c}{1.4cm} w{c}{1.8cm} w{c}{1.4cm} w{c}{1.5cm} w{c}{3.1cm}  w{c}{1.2cm} w{c}{2.5cm} } 
\hline
\textbf{Ref} & Material & Process & Pump & Power & Pair rate & CAR & Fidelity\\
 & & && (mW) & (s$^{-1}$ 100 GHz$^{-1}$) & & \\
\hline
Ref\cite{du2024demonstration} & Si& SpFWM  & CW & 20  & 1.8$\times$ 10$^7$ &   111 & 97.9$\%$\\
&  &   &  & 1  & 3.1$\times$ 10$^5$ &   3613 & 99.85$\%$\\
Ref\cite{olislager2013silicon} & Si& SpFWM  & CW &1.75&   & 8 & 71$\%$\\
Ref\cite{suo2015generation} & Si & SpFWM  & CW & 1.6 &   & 300 & 96.3$\%$\\
Ref\cite{li2017chip} & Si& SpFWM  & CW & 1.37 & 4.6$\times$ 10$^5$  & 75 & 93.4$\%$\\
Ref\cite{miloshevsky2024cmos} & Si& SpFWM  & CW & 6.3 &   &  & 94$\%$\\
Ref\cite{matsuda2012monolithically} & Si& SpFWM  & Pulse & 69(p) & 6$\times$ 10$^5$  & 55 & 91$\%$\\
Ref\cite{takesue2008generation} & Si& SpFWM  & Pulse & 60(p) &   &  & 94.8$\%$\\
Ref\cite{zhang2019generation} & Si& SpFWM  & Pulse & 0.12(a) & 2.7$\times$ 10$^5$  & 230 & 95$\%$\\
Ref\cite{zhang2024polarization} & SiN& SpFWM  & CW & 2.24 & & 82 & 75.7$\%$\\
Ref\cite{wen2023polarization} & SiN& SpFWM  & CW & 2.76 & 1.4$\times$ 10$^5$  & & 81.5$\%$-96$\%$\\
Ref\cite{kultavewuti2017polarization} & AlGaAs& SpFWM  & Pulse & & 1.64$\times$ 10$^5$  & & 90$\%$\\
Ref\cite{orieux2013direct} & AlGaAs& SPDC  & Pulse & & 7$\times$ 10$^2$  & 19 & 83$\%$\\
Ref\cite{kim2025integrated} & TFLN & SPDC  & CW &  &  &  & 94.4$\%$\\
Ref\cite{jiao2025electrically} & TFLN & SPDC  & CW & 0.0188 & 9.3 $\times$ 10$^6$ & 38 & 96.5$\%$-97.1$\%$\\
This work & TFLN & SPDC  & CW & 0.63 & 3.4 $\times$ 10$^7$ & 510 & 99.3$\%$-99.8$\%$\\
\hline
\end{tabular}
\end{center}
\end{table*}

\begin{figure*}[htbp]
\centering 
\includegraphics[width = 6.3in]{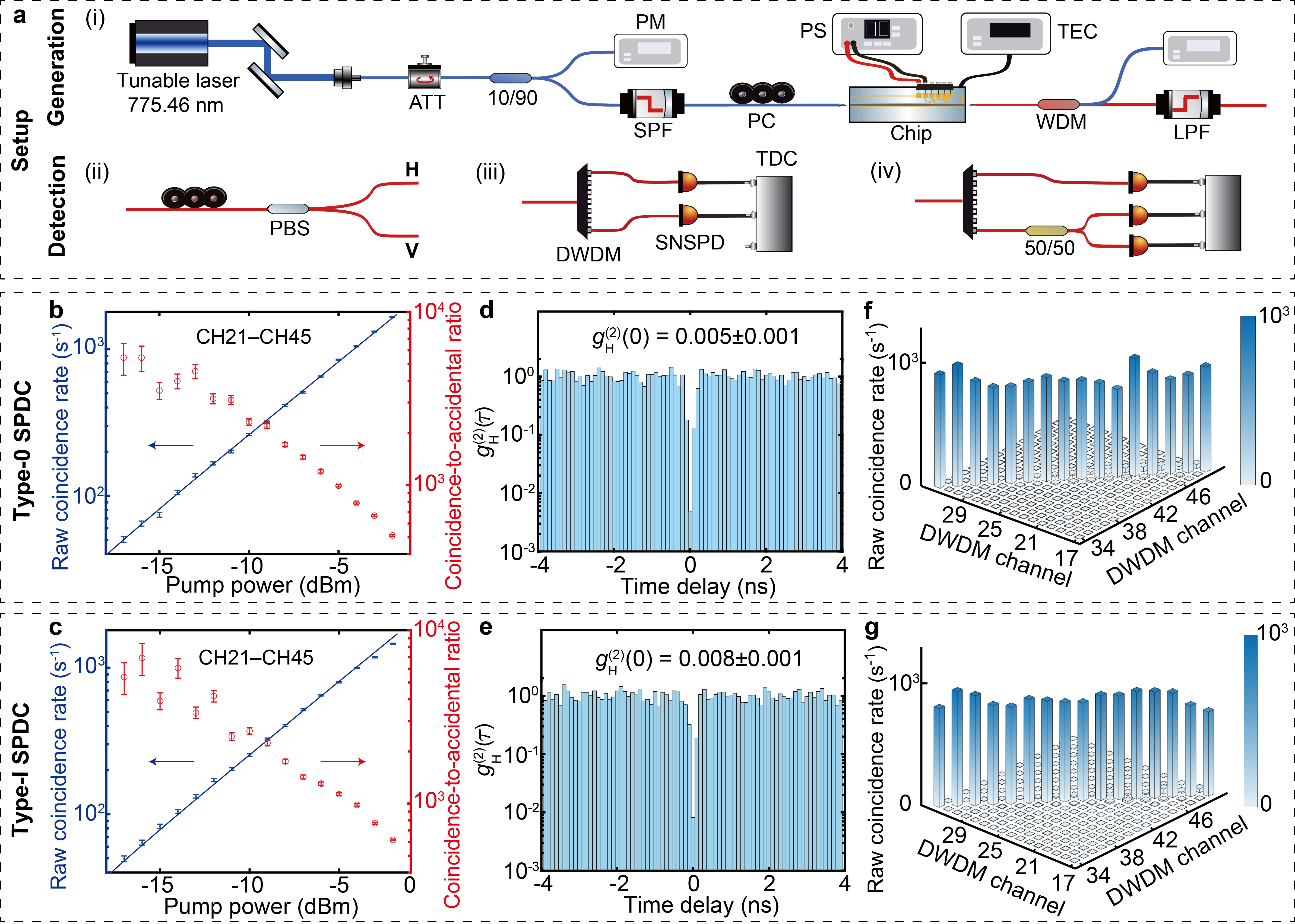}
\caption{\textbf{Non-classical characterization of D-QPM PPLN photon-pair source.} \textbf{a}, Experimental setup for (i) polarization-entangled photon-pair generation, (ii) separation of type-0 (H) and type-I (V) photon pairs, (iii) measurement of wavelength-multiplexed pair generation rate and coincidence-to-accidental ratio (CAR), and (iv) measurement of heralded second-order correlation. ATT: tunable attenuator, 10/90: 10/90 beam splitter; PM: power meter; SPF: short-pass filter; PC: polarization controller; PS: power supply; TEC: thermoelectric cooling; WDM: 1550 nm/775 nm wavelength division multiplexer; LPF: long-pass filter; PBS: polarization beam splitter; DWDM: dense wavelength division multiplexer; SNSPD: superconducting nanowire single-photon detector; TDC: time-to-digital converter; 50/50: 50/50 beam splitter. \textbf{b}, \textbf{c}, Measured raw coincidence counts (blue dots) and quadratic fits (blue line) for type-0 and type-I SPDC, along with CAR (red) as a function of pump power, using selected channel pair CH21-CH45. \textbf{d}, \textbf{e}, Heralded second-order correlation function for type-0 and type-I SPDC at 0.63 mW pump power. Values below 0.01 at zero time delay confirm single-photon operation. \textbf{f}, \textbf{g}, Joint spectral intensity of type-0 and type-I SPDC, reconstructed from correlations across 32 DWDM channels (CH17 to CH32 and CH49 to CH34). The results demonstrate broadband operation with strong frequency correlations.}
\label{Fig3}
\end{figure*}

\begin{figure*}[htbp]
\centering 
\includegraphics[width = 5.8in]{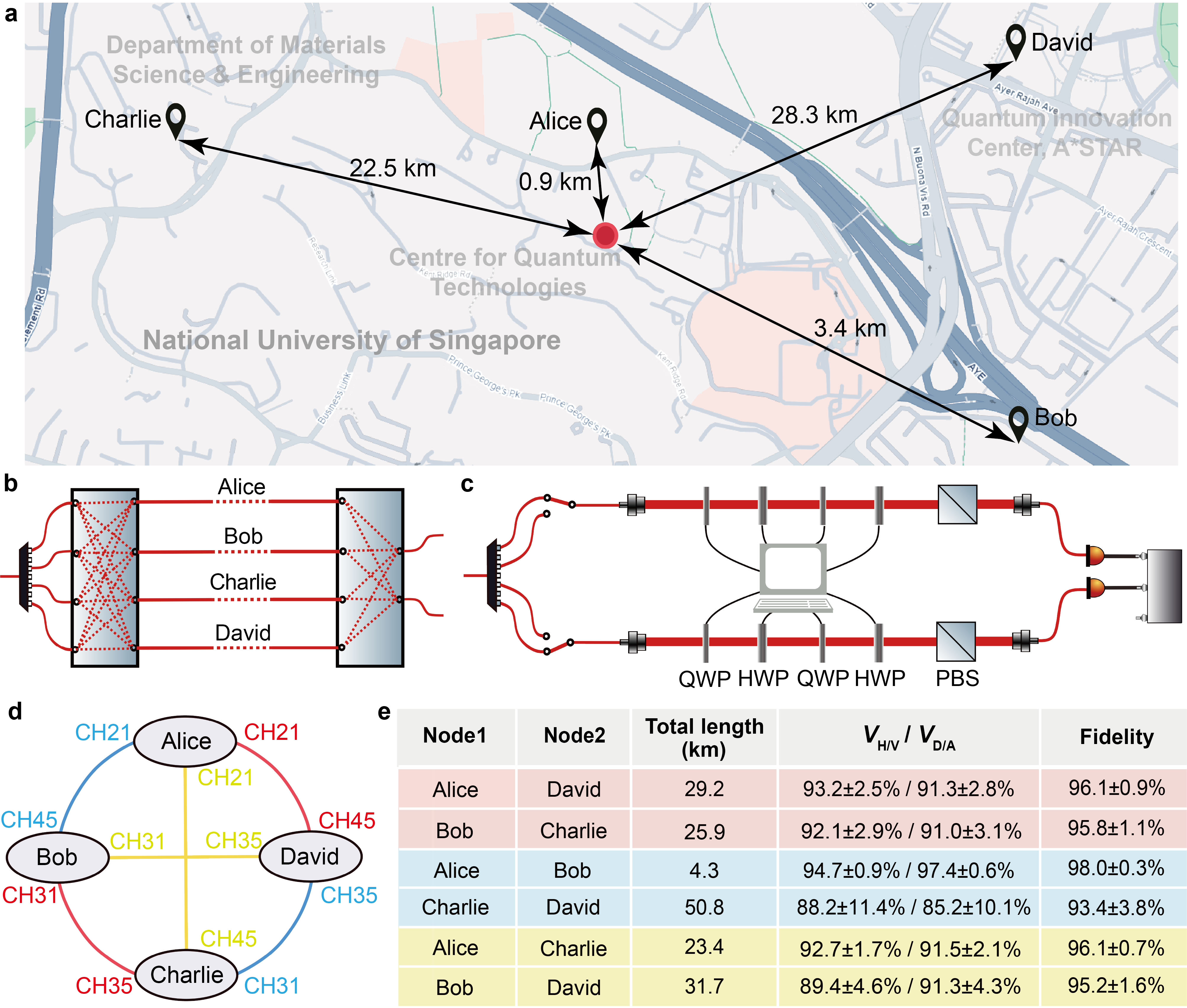}
\caption{\textbf{Long-distance multi-user entanglement demonstration over deployed metropolitan optical fibers.} \textbf{a}, Map of the deployed optical fiber network used to realize a four-user wavelength-multiplexed quantum network (Map data from Google Earth). 
\textbf{b}, Schematic of network operation, where entangled photons from two DWDM channel pairs are distributed to four nodes (Alice, Bob, Charlie, and David), and looped back for detection. The loopback deployed fiber lengths are 0.9 km, 3.4 km, 28.3 km, and 22.5 km, respectively. Note that these lengths differ from the geographical distances shown on the map due to the practical routing of the deployed fiber network.
\textbf{c}, Experimental setup for polarization-entanglement measurements. QWP: quarter-wave plate; HWP: half-wave plate. 
\textbf{d}, Communication layer of the quantum mesh network. Each user is fully connected to all others via three wavelength-distribution configurations (red, blue, and yellow) using two DWDM channel pairs (CH21-CH45 and CH31-CH35). \textbf{e}, Measured raw visibilities for all two-user connections without accidental subtraction, demonstrating robust polarization entanglement across long-distance fiber links up to 50 km.}
\label{Fig4}
\end{figure*}

\end{document}